\documentclass[rnote]{aa}
\usepackage{txfonts}
\usepackage{longtable}  
\usepackage{rotating}
\usepackage{natbib}
\usepackage{graphicx}
\usepackage{graphics}
\usepackage{psfrag}
\usepackage{amssymb}
\bibpunct{(}{)}{;}{a}{}{,}
\def\Teff{$T_{\mathrm{eff}}$}
\def\logg{\ensuremath{\log g}}
\def\vmic{$\upsilon_{\mathrm{mic}}$}

\def\vsini{\ensuremath{{\upsilon}\sin i}}
\def\kms{$\mathrm{km\,s}^{-1}$}

\def\vr{${\upsilon}_{\mathrm{r}}$}
\def\logt{\ensuremath{\log t}}

\def\llm{{\sc LLmodels}}

\def\logl{\ensuremath{\log L/L_{\odot}}}

\def\errvr{$\sigma_{{\upsilon}_{\mathrm{r}}}$}
\def\asec{\hbox{"\hskip-3pt .}}
\def\R{\ensuremath{R/R_{\odot}}}
\def\M{\ensuremath{M/M_{\odot}}}
\begin{document} 
\title{Explaining the Praesepe blue straggler HD~73666} 
\subtitle{} 
\author{L. Fossati\inst{1,2}       \and
	S. Mochnacki\inst{3}	 \and 
	J. Landstreet\inst{4}    \and 
	W. Weiss\inst{1}         
}
\institute{
	Institut f\"ur Astronomie, Universit\"{a}t Wien, 
	T\"{u}rkenschanzstrasse 17, 1180 Wien, Austria\\
	\email{weiss@astro.univie.ac.at} 
	\and
	Department of Physics and Astronomy, The Open University, Milton Keynes,
	MK7 6AA, UK\\
	\email{l.fossati@open.ac.uk}
	\and
	Department of Astronomy \& Astrophysics, University of Toronto,
	50 St.George St, Rm.101, Toronto,ON, Canada M5S 3H4\\ 
	\email{stefan@astro.utoronto.ca} 
	\and
	Department of Physics \& Astronomy, University of Western Ontario, 
	London, ON, Canada N6A 3K7\\
	\email{jlandstr@astro.uwo.ca}
} 
\date{} 
\abstract 
{}
{The blue straggler phenomenon is not yet well explained by current
theory; however, evolutionary models of star clusters call for a good 
knowledge of it. Here we try to understand the possible formation scenario of 
HD~73666, a blue straggler member of the Praesepe cluster.} 
{We compile the known physical properties of HD~73666 found in the 
literature, focusing in particular on possible binarity and the abundance 
pattern.} 
{HD~73666 appears to be slowly rotating, has no detectable magnetic field, and 
has normal abundances, thereby excluding close binary evolution and mass
transfer processes. 
There is no evidence of a hot radiation source.}
{With the use of theoretical results on blue straggler formation present in
literature, we are able to conclude that HD~73666 was probably formed by
physical collision involving at least one binary system, between 5 and 
350\,Myr (50\,Myr if the star is an intrinsic slow rotator) ago.}

\keywords{}
\titlerunning{Explaining the Praesepe blue straggler HD~73666}
\authorrunning{L.~Fossati et al.}
\maketitle
\section{Introduction}\label{introduction}
Blue stragglers are stars that lay on the extension of the main sequence 
and are bluer and brighter compared to the main sequence turn-off stars.
These objects are found in star clusters, dwarf galaxies and in the 
field. The existence of blue stragglers probably can be explained only by 
an interaction between two or more stars, and so to understand this 
phenomenon we study the interaction of stars in stellar systems 
\citep[see][]{leonard,bailyn95,sandquist97,sills97,chen2004,ahumada2007}.

Recently, \citet{ahumada2007} listed the most frequently cited theories to 
explain the formation of blue stragglers. They could be (1) horizontal-branch 
stars that appear above the main sequence turn-off point, (2) stars of 
second or third generation, (3) stars that have extended their main-sequence 
life due to some internal mixing (this would generate a chemically peculiar 
blue straggler), (4) stars formed by collision of two single stars, (5) the 
result of mass transfer in a close binary system, (6) produced by merger of 
the components of a binary system or (7) the result of collision between two 
or more binary systems. Of those seven theories they considered the last four 
as major channels of formation.

Recently, comprehensive catalogues of blue stragglers in open and globular 
clusters have been published. \citet{ahumada1995} created  the
first consistent catalogue of blue stragglers in open clusters, which was then
expanded by \citet{demarchi} and finally superseded by \citet{ahumada2007}.
These catalogues make it possible to analyse blue stragglers on a 
solid statistical base, leading, for example, to the conclusion that 
the number of blue stragglers increases with cluster age 
\citep[see Fig.~5 by ][]{ahumada2007}. In 
the light of this result, evolutionary models of open clusters need to 
consider this not well understood phenomenon. For this reason it is 
important to find clues which allow us to distinguish between different 
blue straggler formation channels.

The A1V star HD~73666 (40 Cnc, $V$ = 6.61) is an extreme blue
straggler \citep{ahumada2007} in the nearby, well-studied cluster Praesepe 
(NGC~2632). This paper will show that it is a particularly important 
example, and provides an excellent test for theories of blue straggler 
formation. This paper can also be seen as the continuation of 
\citet{conti1974} in the light of more than thirty years of new astronomical 
knowledge and instrumental development.

In Sect.~\ref{sec_membership} we discuss the membership of HD~73666 in the
Praesepe cluster. Fundamental parameters and other physical properties are given
in Sect.~\ref{sec_parameters}, where we show that the star is a blue
straggler. In Sect.~\ref{formation} we describe the possible formation 
scenario. Conclusions are gathered in Sect.~\ref{sec_discussion}.
\section{Cluster membership}\label{sec_membership}
Determination of cluster membership is the first step to demonstrate that a
star is a blue straggler. For this purpose the Praesepe cluster is a perfect
object since it has proper motions and mean radial velocity that are
distinct from the field stars in its vicinity. This peculiarity can be
seen on the atlas of the Praesepe cluster published by \citet{kharchenko2005}
where they show the proper motions of the cluster members, concentrated around
the mean ($-$35.90, $-$12.88)\,mas/yr, compared with the motions of the field 
stars lying in the same region of ($-$8.68, $-$1.37)\,mas/yr with a dispersion,
of the cluster stars, of $\sigma_{\mu_{AF}} = 13.58 \pm 3.71$\,mas/yr 
\citep{kharchenko2004}. 

In recent years  major studies of cluster membership have been
published by \citet{chen}, \citet{robichon1999}, \citet{baumgardt}, 
\citet{kharchenko2004} and \citet{dias2006}. HD~73666 is considered by 
\citet{robichon1999} and \citet{baumgardt} to be a cluster member, and used to 
derive the mean cluster astrometric parameters and mean radial velocity. 
\citet{kharchenko2004} as well consider HD~73666 as a cluster member 
having a kinematic membership probability of 0.3175, a photometric membership 
probability of 1.0000 and a positional probability of 1 (Kharchenko considers 
every star with a kinematic and photometric membership probability higher 
than 0.14 as a probable member of the cluster).

Table~\ref{velocity} shows a comparison between the cluster mean
values of parallax, proper motion and radial velocity for Praesepe, and
the individual values for HD~73666.  $\mu_{\alpha}\cos(\delta)$ and
the radial velocity are in agreement within one $\sigma$, but
$\mu_{\delta}$ for the cluster and the star differ by $2.7 \sigma$.
If one takes the cluster members listed by \citet{robichon1999}, and 
selects from the recent re-reduction of the Hipparcos data by 
\citet{vanLeeuwen} the 12 proper motions with measurement uncertainties 
comparable to those of HD~73666, it is found that the dispersion of 
$\mu_{\delta}$ is approximately 1.1\,mas/yr, a value very similar to
the difference between the Praesepe $\mu_{\delta}$ cluster mean and
the value for HD~73666. Thus the apparent discrepancy between cluster 
and star values of $\mu_{\delta}$ is typical of all the most precise 
$\mu_{\delta}$ values for the cluster. The origin of this
effect is not clear to us, but our conclusion is that the agreement
between cluster and stellar mean values is as satisfactory for
HD~73666 as it is for most of the most precisely measured stars in
the cluster, and is not a cause for concern for this specific star.

\begin{table*}[ht]
\caption[ ]{Proper motion and radial velocity of the Praesepe cluster and HD~73666.}
\label{velocity}
\begin{center}
\begin{tabular}{ccccccccccccc}
\hline
\hline
 & RA & DEC & $\pi$ & $\sigma_{\pi}$ & $\mu_{\alpha}\cos(\delta)$ & $\sigma_{\mu_{\alpha}\cos(\delta)}$ & $\mu_{\delta}$ & $\sigma_{\mu_{\delta}}$ & \vr    & \errvr & \Teff & \logl\, \\
 &    &     & [mas] & [mas]          & [mas/yr]                 & [mas/yr]                              & [mas/yr]       & [mas/yr]                & [\kms] & [\kms] & [K]   & [dex] \\
\hline											
Praesepe & 08 40.4 & +19 41 & 5.49 & 0.19 & $-$35.68 & 0.30 & $-$12.72 & 0.25 & 34.5 & 0.0 & (...) & (...) \\
HD~73666 & 08 40.2 & +19 58 & 5.53 & 0.50 & $-$35.52 & 0.62 & $-$13.97 & 0.39 & 34.1 & 0.4 & 9380  & 1.71  \\
\hline
\end{tabular}
\end{center}
\smallskip 

\vr, \errvr\, and \Teff\, of HD~73666 are taken from \citet{luca1}. \logl\, of HD~73666 is taken from \citet{luca2}. All the other data are taken from \citet{vanLeeuwen}.
\end{table*}


\begin{figure}[ht]
\begin{center}
\includegraphics[width=90mm]{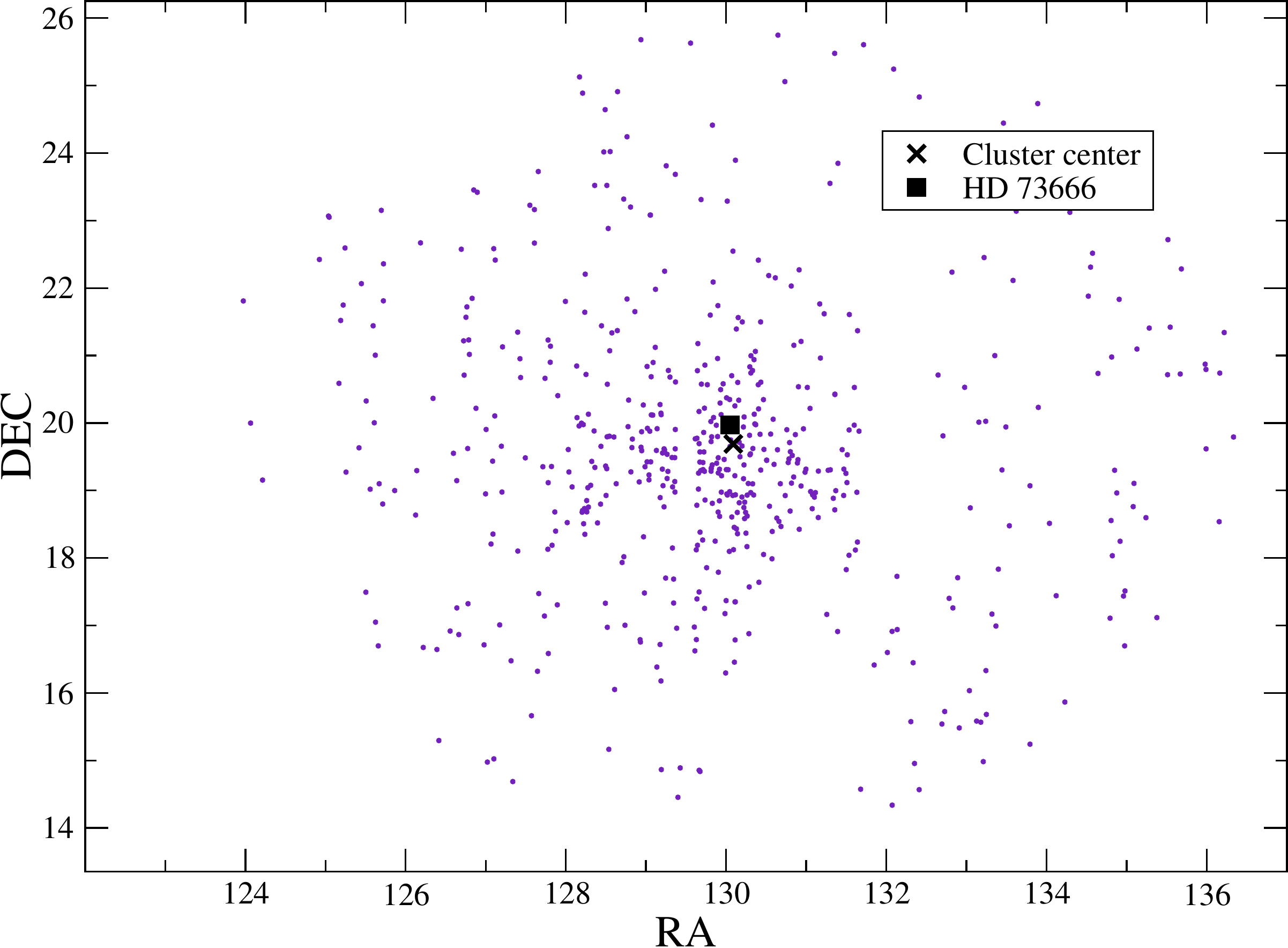}
\caption{Location of HD~73666 relative to stars with a membership probability 
higher than 40\% as tabulated in \citet{adams}. The center of the cluster is
also shown.}
\label{map}
\end{center}
\end{figure}

Figure~\ref{map} shows the projected location of HD~73666 relative to other 
cluster members, which is very near to the center of the cluster. In 
conclusion, there is very little reason to doubt that HD~73666 is a member of 
Praesepe.
\section{Fundamental parameters and Color-Magnitude diagram}
\label{sec_parameters}
\begin{figure}[ht]
\begin{center}
\includegraphics[width=90mm]{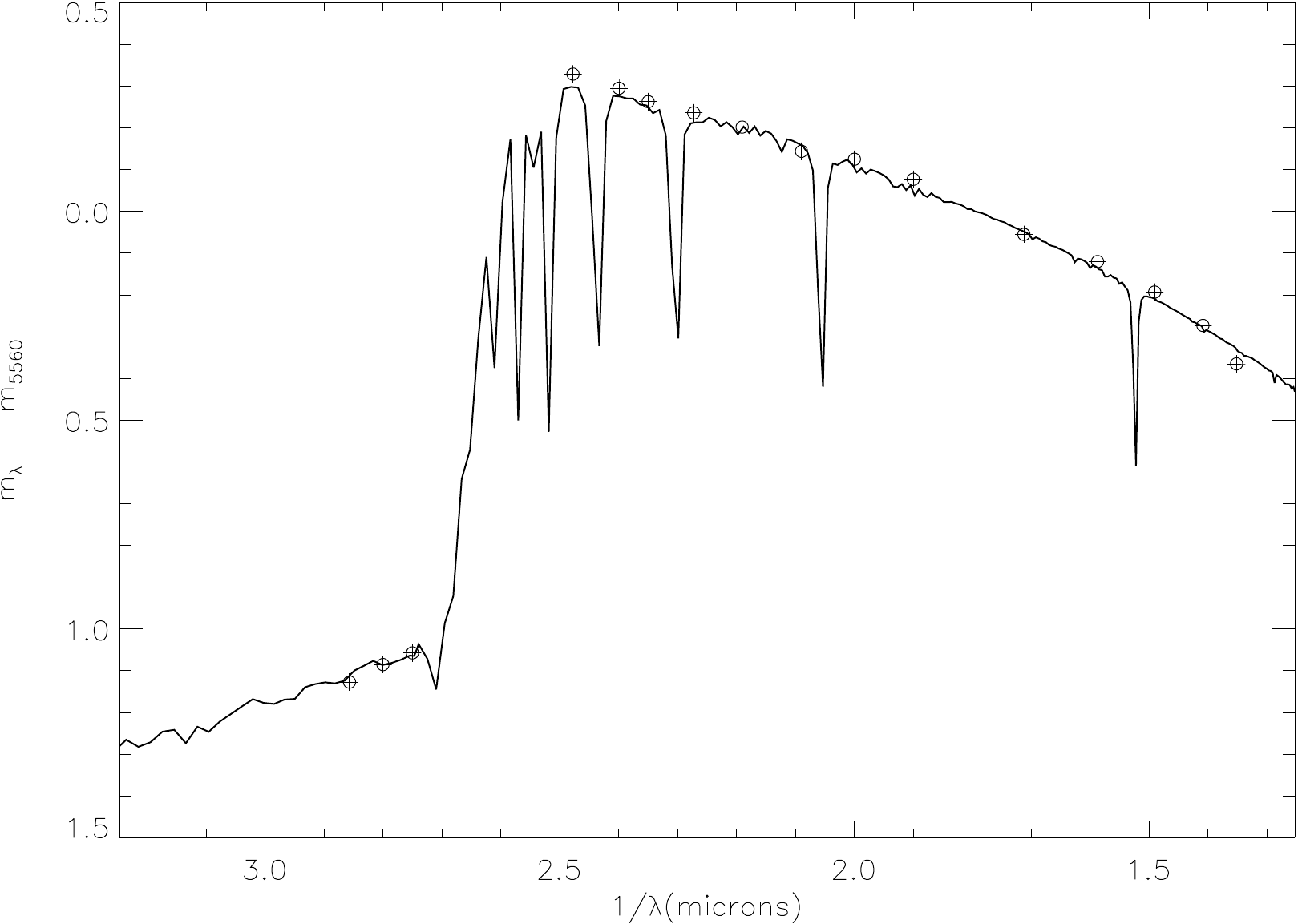}
\caption{Comparison between spectrophotometry from \citet{clampitt} and
theoretical fluxes normalised at 5560~\AA\, with the fundamental parameters 
of HD~73666.} 
\label{sph} 
\end{center} 
\end{figure}

\citet{luca1} derived, from high resolution spectroscopy, \Teff\ and \logg\ of
HD~73666. They obtained \Teff\ = 9380$\pm$200\,K and \logg\ = 3.78$\pm$0.2. 
These parameters are confirmed also by spectrophotometry taken from 
\citet{clampitt}: we calculated theoretical stellar fluxes with the model
atmosphere code ATLAS9 \citep{kurucz1979} with these values of \Teff\ 
and \logg\ and then normalised the fluxes at 5560\,\AA. The comparison between 
the spectrophotometry and the normalised fluxes is shown in Fig.~\ref{sph}.

\begin{figure}[ht]
\begin{center}
\includegraphics[width=90mm]{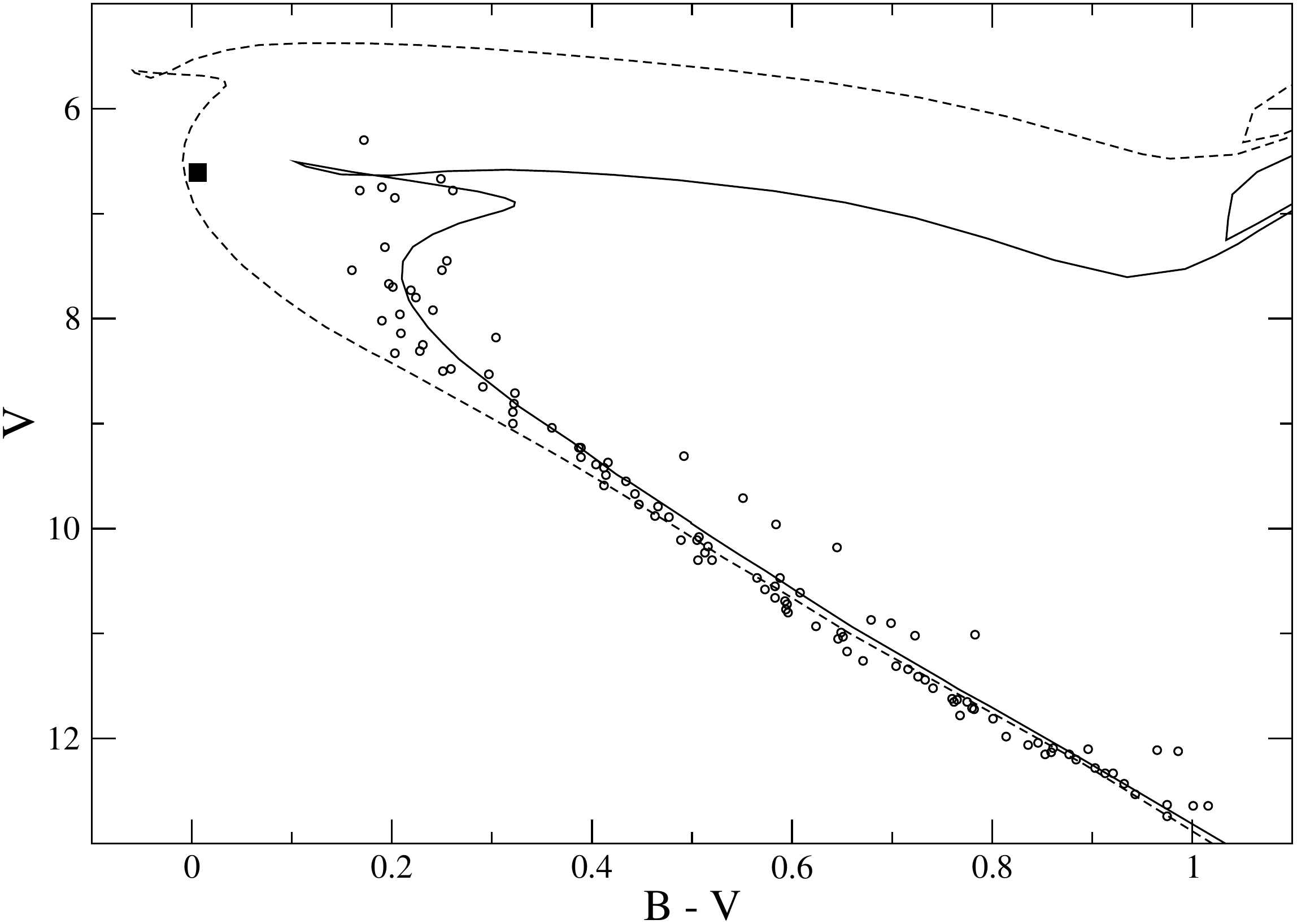}
\caption{Color-magnitude diagram of the Praesepe cluster. The full 
square indicates the position of HD~73666. The photometry was taken from 
\citet{johnson}. The full line shows an isochrone from \citet{girardi2002} for 
the age and metallicity given in the literature by \citet{gonzalez} and 
\citet{chen2003} respectively (\logt\ = 8.85\,dex; 
Z = 0.024). The dashed line represents an isochrone with the same 
metallicity, but an age of \logt\ = 8.55, that best fits the position of 
HD~73666 in the HR diagram.}
\label{hr} 
\end{center} 
\end{figure}

Figure~\ref{hr} shows the color-magnitude (CM) diagram of the cluster,
built using the photometry taken from \citet{johnson}. In the plot we display 
also two isochrones from \citet{girardi2002}, with metallicity 
Z = 0.024$\pm$0.002\,dex taken from \citet{chen2003}, but with different ages. 
The full line represents the isochrone corresponding to the cluster age of 
\logt\ = 8.85$\pm$0.15\,dex \citep{gonzalez}, while the dashed line represents 
the isochrone that best fits the position of HD~73666 on the HR diagram, 
corresponding to an age of \logt\ = 8.55. In Fig.~\ref{hr} we adopted a 
distance modulus of 6.30\,mag \citep{vanLeeuwen} and a reddening of 0.009\,mag
\citep{webda,vdB}.

The CM diagram shows clearly the blue straggler property of HD~73666. 
According to \citet{luca2}, HD~73666 appears to be about 1200\,K hotter 
than the hottest main sequence star member
of the cluster. The mass of HD~73666 that \citet{luca2} derived, from the
Hertzsprung-Russell (HR) diagram, is 2.46$\pm$0.12\,M$_{\odot}$. This value is
about 0.4\,M$_{\odot}$ higher than the mass of the other most massive (mass 
higher or equal to 2\,M$_{\odot}$) main sequence cluster stars. Infact 
following the results of \citet{luca2}, the four most massive main sequence 
stars (excluding HD~73666) are: 
HD~73618 (2.16$\pm$0.22\,M$_{\odot}$), 
HD~72846 (2.09$\pm$0.15\,M$_{\odot}$), 
HD~73711 (2.08$\pm$0.10\,M$_{\odot}$), and 
HD~73709 (2.00$\pm$0.14\,M$_{\odot}$).
From \Teff\ and \logl\ we derive a radius of \R\ = 2.72$\pm$0.12.

One of the possible explanations of the existence of blue stragglers is that 
the star is a horizontal branch star with the same temperature and luminosity 
of main sequence stars. This explanation does not seem very probable: 
the horizontal branch is bluer in a low-metallicity environment 
\citep[see e.g.][]{sandage1960} and
the metallicity of the Praesepe cluster is Z = 0.024$\pm$0.002\,dex 
\citep{chen2003} showing that Praesepe is not a low-metallicity open cluster.
However, a few such stars might exist unrecognised, and so we examine this 
possibility.
 
\begin{figure}[ht]
\begin{center}
\includegraphics[width=90mm]{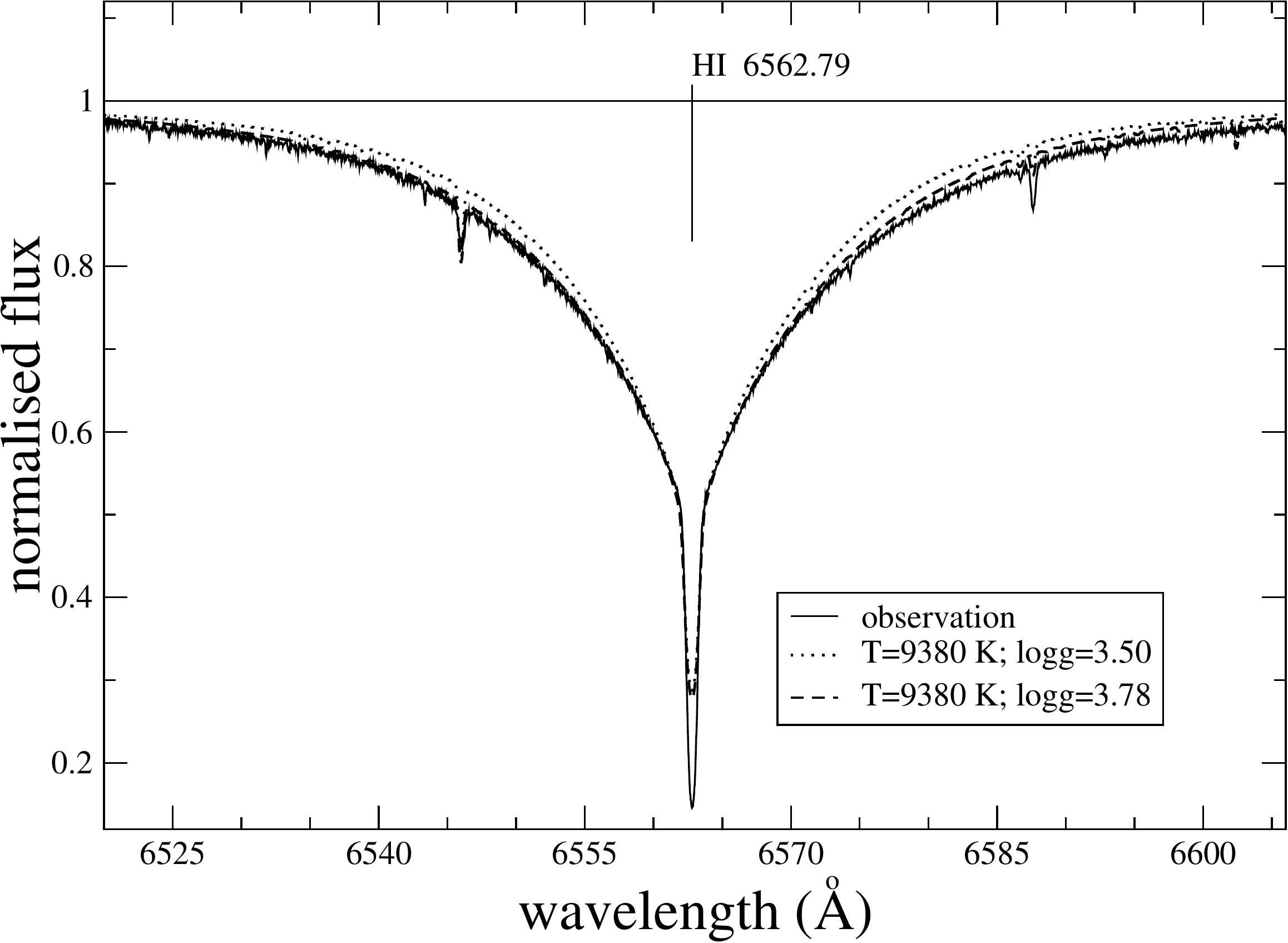}
\caption{Comparison between the observed and synthetic H$\alpha$ line profile 
assuming a fixed \Teff\ = 9380\,K and two different \logg: 3.78 
(assumed -- dashed line) and 3.5 (maximum for an horizontal branch star --
dotted line).} 
\label{halpha} 
\end{center} 
\end{figure}

What distinguishes horizontal branch from main sequence stars is their
mass-radius relation leading to \logg\ values different from the ones 
typical of main sequence stars. 
For horizontal branch stars at \Teff\ $\sim$ 9000\,K \logg\ ranges between 
3.1 and 3.5 \citep{hbs}. For a 
star of the same values of \Teff\ and \logl\  (and hence \R) that we observe, 
but a mass of $0.6 M_\odot$, \logg\ would be $3.35 \pm 0.05$. 
Figure~\ref{halpha} shows the observed H$\alpha$ line profile of HD~73666 in 
comparison with two synthetic profiles, calculated with Synth3 \citep{synth3}, 
assuming \Teff\ = 9380\,K, \logg\ = 3.78 (dashed line) and 
\logg\ = 3.5 (dotted line). We calculated the probability that the observed
H$\alpha$ line profile is fitted by the synthetic profile calculated with 
\logg\ = 3.78 or \logg\ = 3.5. With the higher \logg\ we obtained a 
probability of 99.95\%, while with the lower \logg\ the probability is of 
0.05\%.
The lower \logg\ value is too low, indicating that HD~73666 is very probably 
not a horizontal branch star appearing above the turn-off point. This 
statement disagrees with the conclusion of \citet{conti1974}, although their 
\logg\ determination (\logg\ = 3.7$\pm$0.1\,dex) agrees quite well with the 
adopted one of \citet{luca1}. The conclusion of \citet{conti1974} that
HD~73666 is most likely an horizontal branch star is based just upon 
considerations about stellar evolution and precisely on the fact that Praesepe 
could statistically host one horizontal branch star. We cannot completely 
exclude the possibility that HD~73666 is a horizontal branch star, but we 
conclude that this is not very probable. 
\section{Abundances}\label{sec_abundances}
\begin{figure}[ht]
\begin{center}
\includegraphics[width=90mm]{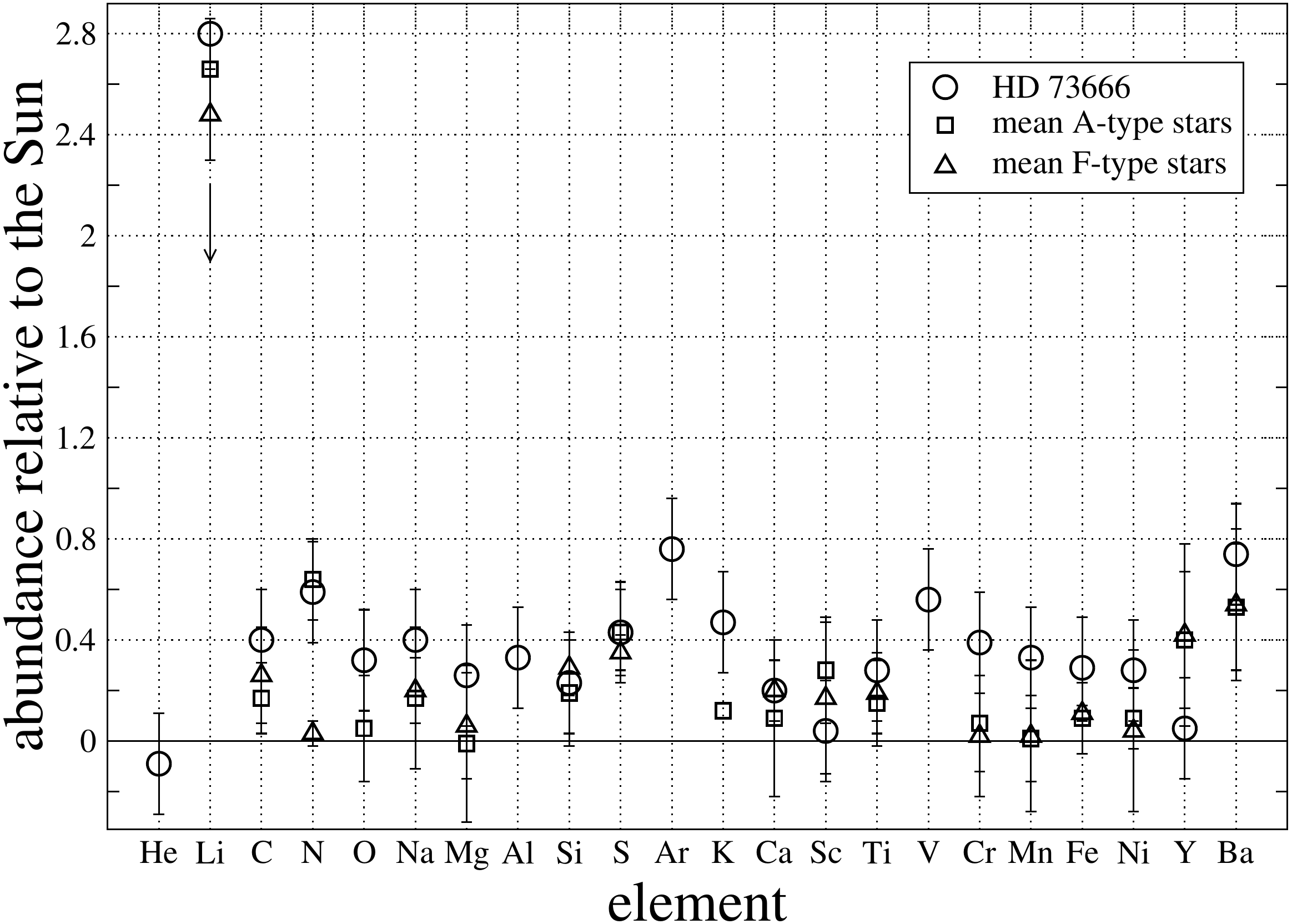}
\caption{Comparison between the abundances of the analysed elements 
obtained for HD~73666 (open circles) and the mean abundances obtained for the 
A- (open squares) and F-type (open triangles) stars member of the cluster. The 
error bar given to the abundances of HD~73666 is fixed to 0.2\,dex for 
all elements. The uncertainty given to the mean abundances are the 
standard deviations from the mean abundance. The arrow indicates that the 
Li abundance given for HD~73666 is an upper limit. All the given abundances 
are relative to the Sun \citep{met05} and were taken from \citet{luca1} 
and \citet{luca2}.} 
\label{abundance} 
\end{center} 
\end{figure}

The abundance pattern of HD~73666 was recently derived by \citet{andri1998}, 
\citet{burkcoupry1998} and \citet{luca1}. The three abundance patterns are all
comparable; the most complete determination is given by \citet{luca1} 
and is reproduced in Fig.~\ref{abundance}, together with the mean abundance 
pattern of the A- and F-type stars member of the cluster\footnote{The 
abundances are in 
$\log(N_{X}/N_{\rm tot})-\log(N_{X}/N_{\rm tot})_{\rm solar}$.}. 

The main characteristics of the abundance pattern of HD~73666 
are: a solar He abundance, overabundance of about 0.4\,dex for C, N and 
O, solar abundance for Ca and Sc (that excludes an Am classification) and a 
mean overabundance of about 0.4\,dex for the other analysed elements (Na, Mg, 
Al, Si, S, Ar, K, Ti, V, Cr, Mn, Fe, Ni, Y and Ba). The Li abundance shown in
Fig.~\ref{abundance} (+2.8\,dex, relative to the Sun) is an upper limit since
at the \Teff\ of HD~73666 the strongest Li line visible in the optical region 
(at $\lambda$ $\sim$6707\,\AA) appears in a synthetic spectrum with a Li 
overabundance of +2.8\,dex (relative to the Sun), and there is no trace of 
this line in the observed spectrum. The Li overabundance obtained for six Am 
stars around the turn-off point is about 2.1\,dex \citep{burkcoupry1998}, too 
low to be detected in HD~73666, in any spectral region. Thus we do not know 
if the Li abundance is lower, similar to, or higher than the one of the 
turn-off stars.

With the use of the Least Square Deconvolution (LSD) technique 
\citep{donatietal1997,wade2000}, applied to Stokes~$V$ spectra, \citet{luca1} 
also searched for the presence of a magnetic field. The measured
longitudinal magnetic field was of 6$\pm$5\,G, showing clearly that the star is
not magnetic.
\section{Other formation scenarios}\label{formation}
\subsection{HD~73666 as a second generation star}
One possible formation scenario for a blue straggler is that the object 
is a second or even third generation star. Multiple episodes of star formation
manifest themselvs in a split of different evolutionary sequences when observed
in a cluster color-magnitude diagram. Multiple stellar populations were
already discovered in galactic and Magellanic Cloud clusters \citep{piotto}.

The color-magnitude diagram of the Praesepe cluster, displayed in 
Fig.~\ref{hr}, shows clearly the presence of one evolutionary sequence. If a
second evolutionary sequence is present, the further episode of star formation
happened at a time within the uncertainty given for the cluster age. The 
possibility that HD~73666 is a second or third generation star would imply
the existence of a star formation mechanism that is able to form a single star
of about 2.5\,\M\ close to the center of the cluster. We find this possibility
extremely unlikely.
\subsection{Mass transfer}\label{sec_binary}
\subsubsection{Distant companion}
\citet{hartkopf1984} observed HD~73666 using speckle interferometry to
determine binarity, because of the overluminosity of the star, but did 
not find a companion. \citet{mcalister} made speckle
interferometry observations of HD~73666 in 1983. They discovered the
presence of a companion star at a separation $\rho$=0\asec425 $\pm$ 0\asec009 
and a position angle $\theta$=127.6 $\pm$ 0.5\,degrees on 1983.0477. 
\citet{mason1993} observed the star another 11 times, detecting  the companion 
only twice, on 1986.8922 ($\rho$=0\asec434 $\pm$ 0\asec009, 
$\theta$=133.7 $\pm$ 0.5\,degrees) and 1987.2664 
($\rho$=0\asec425 $\pm$ 0\asec009, $\theta$=134.1 $\pm$ 0.5\,degrees), 
concluding that the star "shows little orbital motion". These measurements 
suggest that the companion star probably has common proper motion with 
HD~73666. 

The secondary is 2.5$\pm$0.5 magnitudes fainter (Mason, private 
communication) than the primary star. This would make the secondary an F star 
of about 1.5\,M$_{\odot}$.
At a distance of 180\,pc, the angular separation would 
mean a minimum separation at this time about 80\,AU, which suggests a 
period of order 450\,yr or more.
The derived separation of 80\,AU excludes any interaction between the two 
stars, unless the mutual orbit is extremely eccentric, a characteristic which 
would hardly survive a period of heavy mass exchange. 

Binarity has several consequences. Firstly, it favors the present location 
of the system at the cluster center, since Praesepe clearly shows mass 
segregation effects \citep{kraus2007}. Secondly, collisions involving binary 
systems are an effective way of forming blue stragglers \citep{leolin92}.
\subsubsection{Close companion}
The detection of a white dwarf or subdwarf companion close to HD~73666 would 
be very important, as it would clearly show that the blue straggler 
has been formed by mass transfer. Such a companion could be detectable 
through variable radial velocity or visible in the spectrum of HD~73666. 

The presence of a small close companion (white dwarf or subdwarf)
can be tested by searching for radial velocity variability.  \citet{luca2}, 
in their Fig.~13 and Fig.~14, show that the star had no radial velocity 
variations between January 2006 and March 2007. The two measured radial 
velocities are also in agreement (see also Table~\ref{velocity}) with 
the cluster mean. \citet{luca2} also collected the radial velocity 
measurements obtained in the past by \citet{abt1970}, \citet{conti1974}, 
\citet{abt1999} and \citet{soren2002}, concluding that all these measurements 
are consistent with one another and with the cluster mean. In particular 
\citet{conti1974} measured the radial velocity 33 times over 14 years. They 
tried to fit a periodic function to their data and concluded that no 
periodic variation can be fit to the data. From these data they derived the 
probability that a close companion could go undetected, assuming a random 
orientation of the orbital plane (see their Fig.~1). They assumed a 5:1 mass 
ratio (compatible with the possibility of a subdwarf or white dwarf companion) 
and a conservative velocity amplitude $K_1$ $\leq$ 1\,\kms. Note that a 
companion as distant as 15\,AU, about the limit possible for orbital stability 
\citep{bailyn1987} and well beyond the limit for significant mass transfer, 
would have an orbital velocity of order 12\,\kms. \citet{conti1974} concluded 
that "for systems likely to be involved in mass exchange, the chance of 
their being undiscovered is remote indeed". 

We extended the analysis of \citet{conti1974} by carrying out a frequency 
analysis on the basis of all the radial velocity data found in literature (65 
data points), which now span a time base of about 52 years. We have excluded a 
priori all the frequency peaks corresponding to periods of less than 1 year, 
since they should have already been previously detected. Three peaks at 
periods between 20 and 70 years are found, but all of them have a probability 
density (which ranges between 0 and 1) of less than 0.00002, making the 
presence of a close companion very unlikely. Note that \citet{luca1} 
listed HD~73666 as a single-line spectroscopic binary (SB1), an 
incorrect classification which should be ignored.

\begin{figure}[ht]
\begin{center}
\includegraphics[width=90mm]{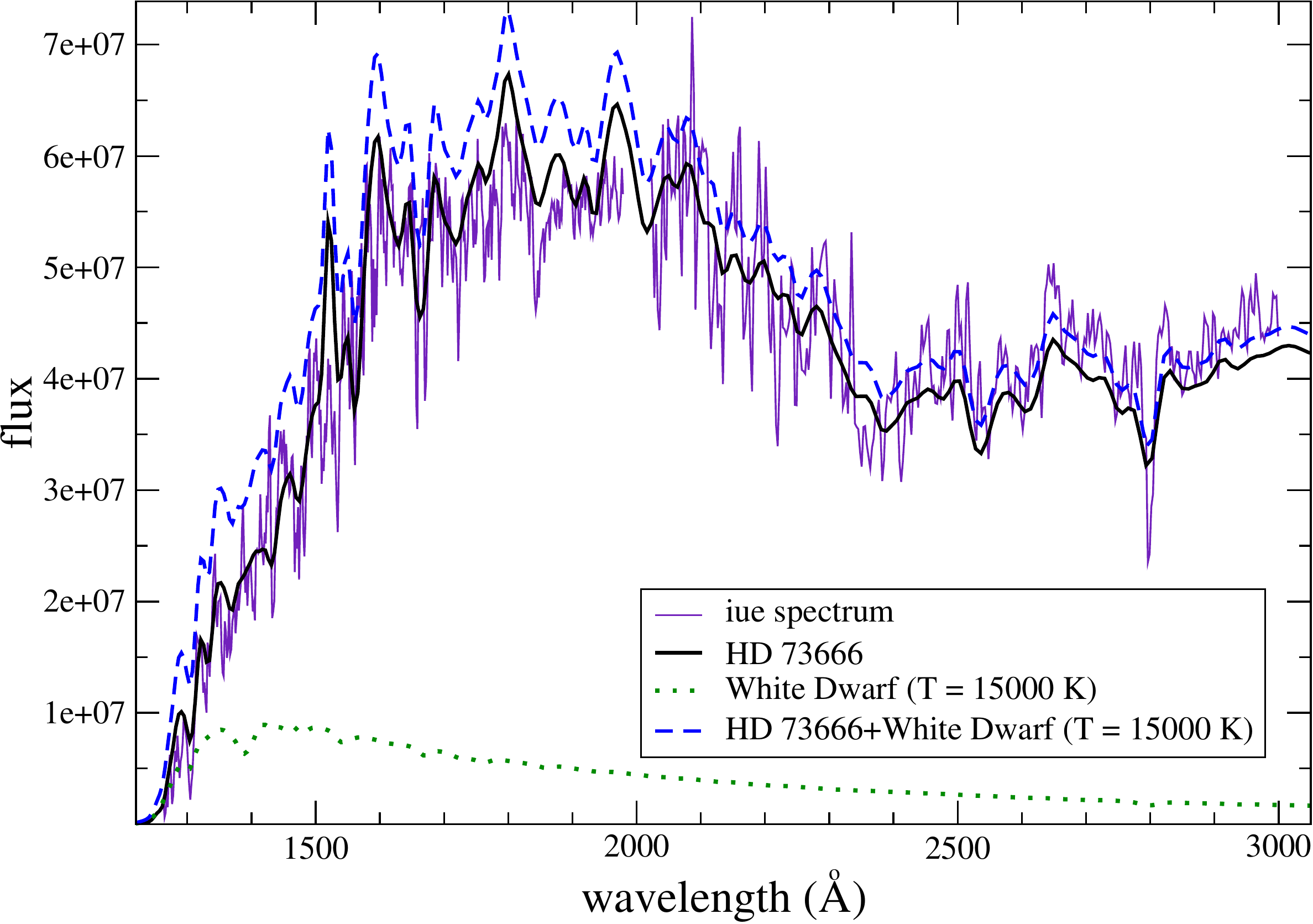}
\includegraphics[width=90mm]{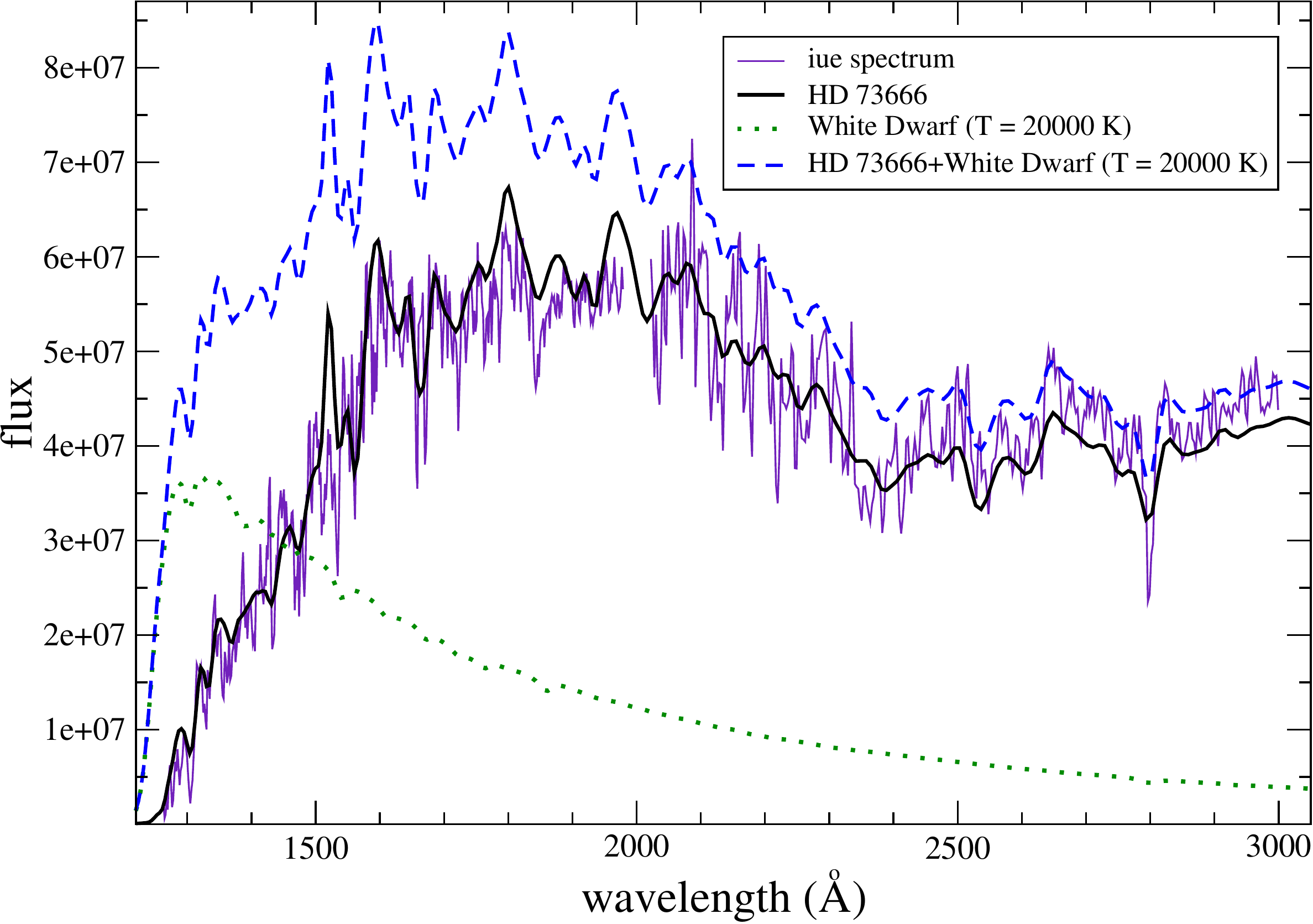}
\caption{Comparison between IUE spectrophotometry (thin full line) and 
theoretical \llm\ fluxes for HD~73666 (thick full line), for a typical white
dwarf (\Teff\ = 15000\,K -- upper panel / \Teff\ = 20000\,K -- lower panel, 
\logg\ = 8.0, \R\ = 0.013, dotted line) and for the two components 
together (dashed line). All the theoretical fluxes take into account the 
estimated stellar radii. The theoretical fluxes have a different resolution 
($\sim$100) than the IUE spectra ($\sim$900) for visualisation reasons.} 
\label{iue} 
\end{center} 
\end{figure}

The presence of a hot close companion can be checked through UV spectra as well.
Figure~\ref{iue} shows a comparison of IUE spectrophotometry of 
HD~73666, calculated fluxes for this star assuming that it is a single 
object, theoretical fluxes for a typical white dwarf (\Teff\ = 15000\,K -- 
upper panel; \Teff\ = 20000\,K -- lower panel),  and 
the sum of the two theoretical fluxes. For the white dwarf we assumed 
\logg\ = 8.0, \R\ = 0.013 and the abundances obtained by \citet{kawka2008} 
for BPM~6502, which has fundamental parameters similar to those assumed here. 
The fluxes were calculated with the use of the LTE code \llm, which uses 
direct sampling of the line opacities \citep{denis2004} and makes it possible 
to compute model atmospheres with an individualised abundance pattern. Taking 
into account the radii of the two objects we have derived the total flux that 
would be visible if HD~73666 were to have this particular white dwarf 
companion. The plot shows that UV fluxes would not clearly reveal the presence 
of a white dwarf companion of \Teff\ = 15000\,K, but would allow us to 
recognise the presence of a hotter white dwarf. 

According to the cooling sequences published by \citet{prada}, a white dwarf 
formed more than half the cluster age ago, would have a luminosity of 
\logl\ $\leq$ $-2$, with a temperature \Teff\ $\leq$ 15000\,K. Thus the 
observed fluxes of HD~73666 do not exclude the presence of a white dwarf 
companion if mass transfer had occurred during the first half of the cluster 
age, within the first 350\,Myr. Figure~\ref{iue} however excludes mass transfer 
onto HD~73666 during roughly the past 200\,Myr. 

On the basis of the analysis of the collected radial velocity measurements and
of the UV analysis we conclude that the primary component of HD~73666 is 
therefore itself very likely a single star, with only the visual companion to 
make it a binary system. If this is not the case and mass transfer produced the
blue straggler, it must have occurred at least of order 200\,Myr ago.
\subsection{Merging and collisional formation scenario}
As mentioned in Sect.~\ref{introduction}, \citet{ahumada2007}  
reviewed the different theories of blue straggler formation. From these 
theories we already concluded that both the horizontal branch confusion, 
multiple episodes of star formation and mass transfer from a close companion are
quite improbable. Among the seven scenario, the remaining possible channels of 
blue straggler formation for HD~73666 are: collisional mergers of two stars, 
merger of the two components of a close binary system, and collisional mergers 
between binary systems.

To consider stellar mergers and collisions as an effective way to form blue
stragglers, it is important to estimate the merger and collision probability. 
\citet{zwart} modelled the evolution of Praesepe-like open clusters, using 
simulations that include stellar dynamics and effects of stellar evolution. 
\citet{zwart} considers that all mergers are due to binaries: either
binary coalescence due to unstable mass transfer, or due to perturbation by 
a third body. They concluded that for Praesepe-like open clusters the 
collision rate is about one per 100\,Myr and the merger rate of two components 
of a close binary system is about one per 50\,Myr. Taking into account an age 
for the Praesepe cluster of about 700\,Myr, the fact that Praesepe shows mass 
segregation \citep{kraus2007} and that HD~73666 is placed close to the cluster 
center, it is highly probable that within the given cluster age stellar 
collisions and mergers happened.

\citet{shetrone} derived abundances of blue stragglers and turn-off stars 
in M~67, to obtain chemical signatures to distinguish between stragglers formed
by collision or by binary mass transfer. They mention that a severe lack of 
lithium could be an important signature of a stellar collision 
\citep{lombardi2002} , although a lack of lithium is expected as well in
blue stragglers formed for mass transfer, so that the lithium depletion alone 
is not enough to indicate a collisional origin. In HD~73666 Li is also not 
observed and it is not possible to detect if this is a temperature effect or 
a real lack of lithium. \citet{shetrone} also concluded that C, N and O may be 
more useful. In particular, a blue straggler formed by collision will not 
change the original CNO abundances, which should be similar to those of the 
turn-off stars. In case of formation by binary mass transfer, CNO should be 
modified and the secondary would become a helium or CO white dwarf. Following 
the model published by \citet{chen2004} it is possible to deduce how much 
the CNO abundances of the primary star would change due to mass transfer. 
According to their models, the oxygen abundance should not vary, while the 
carbon abundance should decrease by about 50\% and the nitrogen abundance 
increase by about 150\%. Both the nitrogen and carbon abundances of 
HD~73666 are comparable with the ones of the other A-type stars of the cluster 
(see Fig.~\ref{abundance}). This result is consistent with our conclusion 
given in Sect.~\ref{sec_binary} that most likely HD~73666 does not have a 
close companion and therefore did not undergo mass transfer. Since we 
have eliminated the alternatives, we conclude that HD~73666 was very probably 
formed by merging or collision.
\section{Discussion and conclusion}\label{sec_discussion}
We next consider whether the surface CNO abundances could provide 
information about a stellar collision. We consider the predictions of 
\citet{sills2005} on the surface helium and CNO abundance of stragglers 
formed after the collision of two low mass stars (0.6\,\M). They found that 
the He and CNO abundance change from the original abundance of the two 
colliding stars but not fast enough to be visible within 350\,Myr, that is 
the estimated maximum age of HD~73666 since becoming a blue straggler 
(see Sect.~\ref{minmax}).

If instead we consider a collision 
between a 2~M$_{\odot}$ star (operating via CNO cycle) and a 0.5~M$_{\odot}$ 
star, we would expect that the CNO present in the core would stay in the core 
of the remnant star, as He does in low mass stars. Note however that the 
collision of a high mass and a low mass star, as forming mechanism for 
HD~73666, is somewhat less probable than the collision of two stars of masses 
near 1\,$M_\odot$, considering that in the center of the cluster the mean 
stellar mass is about 0.8\,M$_{\odot}$ \citep{adams}.
It is also important to mention 
that the models proposed by \citet{lombardi1996} and \citet{lombardi2002} show 
a small mass-loss, during the collision (between 1 and 10\% of the 
total mass of the colliding stars).

We conclude that the abundances of CNO in HD~73666, which are similar to those 
of the turn-off stars, are consistent with collisional formation. 

Since we did not find any evidence to contradict the merging and 
collisional formation scenario, it seems probable that HD~73666 was formed 
through a collisional mergers of two stars or a merger of the two components 
of a close binary system or a collisional mergers between binary systems.

It is important to stress that we are not able to establish which of these
two formation mechanisms is the most probable for HD~73666.
\subsection{Minimum and Maximum Times Since Formation}\label{minmax}
A {\it minimum age} can be deduced if the blue straggler
was formed by a physical stellar collision. \citet{sills97,sills01,sills02} 
showed that the remnants of physical stellar collisions, in
particular of off-axis collisions, should be very fast rotating objects with
typical rotational velocities similar or greater than the break-up velocity.
This brought to the conclusion that if blue stragglers are formed also by 
stellar collisions a mechanism to reduce the angular momentum must exist. 
\citet{sills2005} showed that such spin-down mechanism can be effective 
through mass loss and that the star loses about 80\% of its angular momentum 
within 5\,Myr, while a blue straggler like HD~73666 need about 1.4\,Myr 
to reach the main sequence after the collision. HD~73666 shows an unusual slow 
rotation for an A1V star, so that if the star is actually slowly 
rotating the minimum age of the star as a blue straggler could be set close 
to 5\,Myr. 

\citet{luca1} derived a non-zero microturbulence velocity: 
\vmic\ = 1.9 $\pm$ 0.2\,\kms, but this is a fairly normal value for a star of 
this mass, and appears to indicate the disappearing of large scale 
convection zones, near the surface of the star. This is in agreement with 
several recent modelling of collisionally formed blue stragglers 
\citep[e.g. ][]{sills97,GP08a}.

The {\it maximum age} of HD~73666 since becoming a blue straggler is given by 
the age of the isochrone on which it lies in Fig.~\ref{hr}. This is about 
350\,Myr. If HD~73666 is an intrinsically slow rotator 
\citep[\vsini\ $\leq$ 90\,\kms;][]{charbonneau}, a 
{ \it further maximum age limit} is provided by the time it takes for a
slowly-rotating A1V star to develop chemical abundance anomalies, such as Am
characteristics, by diffusion since the end of convection in its
outer envelope. \citet{talon06} computed evolutionary models with diffusion
for stars of 1.7 to 2.5\,M$_{\odot}$. Their most relevant model, labeled
2.5P2, for a star of mass 2.5\,M$_{\odot}$ and rotation speed 15\,\kms, shows
clear abundance anomalies before 50\,Myr, which we can conservatively set as
the maximum age of HD~73666. This is, of course, valid if HD~73666 is now an 
intrinsically slow rotator. The fact that HD~73666 appears so far off the
ZAMS does not contradict the given maximum age of 50\,Myr since collisionally 
formed blue stragglers appear on the main sequence not on the ZAMS, but at an
already evolved stage.
\subsection{Conclusion}
We conclude that the Praesepe blue straggler, HD~73666, was likely formed by
physical stellar collision and merger of two low-mass stars, between 5 and
350\,Myr ago (50\,Myr if the star is an intrinsic slow rotator) if 
current models are correct.

On the basis of our knowledge of HD~73666 it is not possible to 
distinguish between a direct stellar collision and binary coalescence as the 
formation mechanism for HD73666.

HD~73666 could be a perfect object to test current models of collisionally
formed blue stragglers. The wide and detailed knowledge available about this 
star and the environment in which this star is present would allow to test the
reliability of current models and to give important constraints for their 
future development.
\begin{acknowledgements}
LF has received support from the Austrian Science Foundation 
(FWF project P17890-N2). We thank Dr B. Mason for his help in digging up
interferometric measurements, M. Gruberbauer for the frequency analysis and P.
Reegen for the useful discussions. We are greatful to the anonymous referee 
for the comments and suggestions that improved a lot the manuscript. 
\end{acknowledgements}
\end{document}